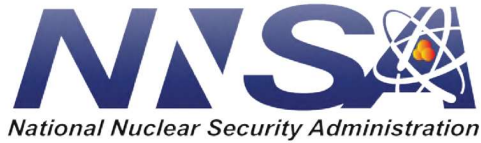 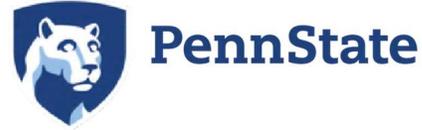

# Comparative Pulse Shape Discrimination and Coincidence Analysis of ($\alpha$,n) Neutron Sources

Nuclear Engineering Nuclear Security Education Laboratory

Term Paper

Sebastian Ritter

April 20th 2020

Nuclear Security Education Laboratory
Radiation Science and Engineering Center

# Contents

# Abstract

The gamma to neutron ratio is measured via Pulse Shape Discrimination (PSD) of a Plutonium Beryllium (PuBe) source and of a rudimentary constructed Polonium Beryllium (PoBe) source using one organic liquid scintillator detector and digital signal processing. Furthermore, two organic liquid scintillator detectors and digital signal processing is used to conduct coincidence counting on the PoBe source. Measured data is compared to literature. The gamma to neutron source ratio was found to be 3.67 ± 0.14 and 1.13 ± 0.05 for the PuBe and for the PoBe source respectively. Both ratios are above literature values for the respective neutron sources. Coincidence counting analysis showed that the PoBe source emitted two or more coincident gamma-rays or neutrons with a confidence level of 91.77 % using a coincident window of 40ns and a detector source distance of 1 mm.

# 1 Introduction

The quantification of bulk samples of uranium or plutonium via active neutron interrogation using organic liquid scintillator detectors is establishing itself as an important tool in nuclear safeguards applications [14]. Different neutron sources exhibit different gamma to neutron emission rates and different neutron source coincidence rates. High neutron source coincidence rates may mask fission neutrons coincidence detections from the interrogated material. Furthermore, high gamma to neutron source-emission rates may saturate detectors and render strong neutron sources impractical for low dose applications. In particular active neutron interrogation in combination with coincidence counting provides for a non-destructive sample analysis technique with acceptable detection limits of special nuclear material (SNM) [3] [7].

Paragraph 7 of INFCIRC/153 Part 1 requires that a state shall "establish and maintain a system of accounting for and control of all nuclear material subject to safeguards", or SSAC for short [2]. INFCIRC/153 requires the IAEA to verify a state's accounting findings and verify a state's non-diversion. The IAEA is required to conduct material verification via independent measurements [2]. In particular IAEA SNM material accounting is known for utilizing neutron coincidence counting [3] [7]. Neutron coincidence counting in passive detection systems has been used in nuclear safeguards applications to determine plutonium and U-235 contents of safeguard relevant objects providing a quantitative verification method [7] [9]. An example of a passive neutron coincidence counting system is portable the Canberra JCC-14 coincidence counter [1].

While the Canberra JCC-14 coincidence counter features a high counting efficiency of 42 %, it utilizes eighteen 3He detectors [1]. With a reported shortage of 3He, alternatives are explored with liquid organic scintillator detectors with digital signal pulse processing being seen by some researchers to be an alternative [5][14][3]. For this reason, this paper utilizes a CAEN liquid organic scintillator detectors with digital signal pulse processing for neutron source analysis.

Two experimental methods are used for neutron source analysis: pulse shape discrimination (PSD) and coincidence counting. PSD is a signal data processing technique that allows for the discrimination of gamma counts against counts which originated from neutron interactions in the detector volume. Coincidence counting is a powerful radiation detection technique with the potential to significantly suppress background detector counts and to identify two or more counts which are closely time related allowing for neutron multiplicity counting. Two or more events are considered coincident if they occur within the same time window. If they do, an event is recorded along with a timestamp without waveform information. Waveform information is only recorded in the PSD analysis method. The coincidence detection system utilized detects both neutron and gamma coincidences which may be any coincidences of two or more in number within a certain coincidence time window.

## 2   Literature Review

A project involving PSD was previously conducted by Scott Wandel at the Pennsylvania State University in May 2014. His report is locally available in the Pennsylvania State University's Academic Projects Building, Room 115. The researcher wrote a pulse shape discrimination (PSD) Matlab code which they made available to this paper. Further this Matlab code is exclusively utilized for PSD analysis in this paper albeit with some modifications in its code to allow for the readout of neutron and gamma counts from the acquired data sets. In addition to the PSD code from Scott Wandel, code from Patrick Egan is utilized which he copyrighted at the Pennsylvania State University in April 2006. In Scott Wandel's computer program, three methods were applied for PSD. The Tail-to-Full Integral Ratio, the Tail-to-Head Integral Ratio and the Tail-to-Pulse Height Ratio. Detailed explanation of these commonly used PSD methods is available in literature [11] [17]. For each PSD method, a figure of merit (FOM) can be defined as the distance between the gamma ray and neutron peaks in units of the PSD parameter divided by the full width at half maximum (FWHM) of the peaks in units of the PSD parameter. A higher FOM is better.

Jennifer L. Dolan, Marek Flaska, Alexis Poitrasson-Riviere, Andreas Enqvist, et al. [5] field tested a fast-neutron multiplicity counter developed at the University of Michigan and characterized 240 Pu effective masses. The neutron coincidence counting system consisted of sixteen 7.62cm diameter by 7.62cm EJ-309 liquid scintillator detectors arranged in a circular pattern placed on to levels of elevation. A CAEN A1536N high-voltage power supply is utilized. The system was calibrated via 252Cf and 137Cs sources as well as well-characterized plutonium samples.

The researchers used two time-synchronized CAEN V1720, 12bit, 250-MHz, 8-channel signal digitizers. The multiplicity detection window in 100ns.The researchers conducted a 2.5 day background measurement yielding 9.17 cps for single neutron detection, 0.037 ± 0.0006 cps for double neutron detection and 0.00017 ± 0.00004 cps for triple neutron detection. These values are noted by the researchers to be higher than expected from the natural background. The PSD false positive single neutron detection rate was estimated to be 2 to 3 in 1000 with a Cs137 source, which is considered a pure mono-energetic gamma emitter. The false positive neutron detection rate for double and triplet neutron events is estimated to be near zero given the narrow 100ns multiplicity detection window. The obtained Cf-252 PSD discrimination line for 200,000 waveforms is found in Figure 1. Using a Cf-252 calibration source the absolute fission detection efficiency of was found to be 17.48 ± 0.33 % for single neutron detection events, 1.10 ± 0.01 for double neutron detection events and 0.047 ± 0.001 for triple neutron detection events. The gamma to neutron ratio for Cf-252 was found to be two. ($\alpha$,n) reactions are seen as a big contributor to single neutron events. No source shielding was used in all quoted information.

N. Ensslin describes [7] neutrons from ($\alpha$,n) reactions to be random and non-coincident making them a sensible interrogation source. Neutrons from spontaneous and induced fission events are coincident and emitted isotropically. Passive neutron coincidence detection systems focus on unique spontaneous fission signatures even in presence of ($\alpha$,n) neutrons. By knowledge of isotope type, isotope quantity can be determined.

A.C.Kaplan et al. found [10] a gamma to neutron ratio for

(a)

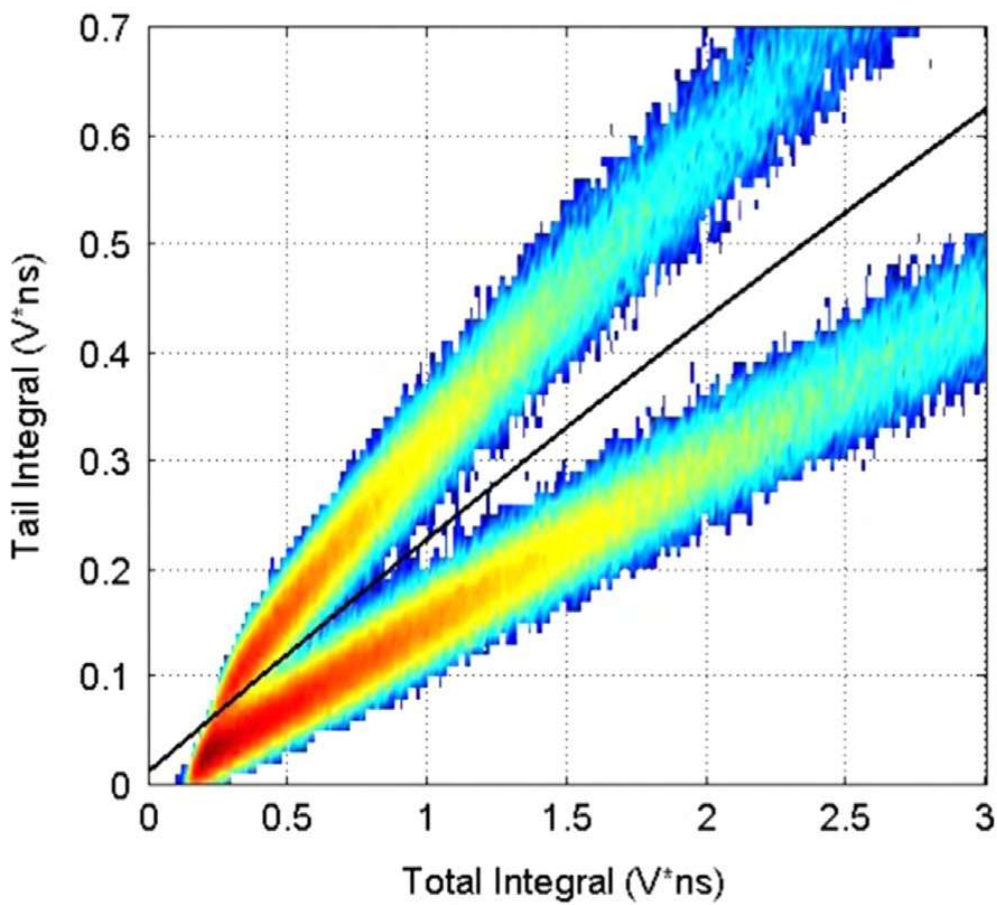

Figure 1: Cf-252 PSD discrimination line for 200,000 waveforms. Credit: Jennifer L. Dolan, Marek Flaska, Alexis Poitrasson-Riviere, Andreas Enqvist, et al. [5].

Cf-252 of 5 using a single 3 by 3 inches EJ-309 organic liquid scintillator detector with an ETL 9821 photo-multiplier tube and a CAEN V1720 eight-channel, 12 bit, 250 MHz pulse signal digitizer.

The gamma and neutron emission energy spectra of PuBe and AmBe sources are well characterized by H´ector Ren´e VegaCarrillo et al. [21] as well as Griffith et al. [8]. H´ector Ren´eVegaCarrillo et al. found a roughly two orders of magnitude increase in the specific gamma to neutron emission ratio for the PuBe source when compared to the AmBe acorss the gamma energy spectrum.

G.Venkataraman et al. [22] calculated the gamma to neutron ratio to be 0.73 for both a PuBe and a AmBe source. The researchers also measured the gamma to neutron ratio using an experimental setup consisting of a 65 mCi PuBe source, a 30 mCi AmBe source, 5 by 4 inches NaI scintillator with a 256 channel multi-channel analyzer (MCA) and a detector source distance of 25cm. They found the ratio to be analyse 0.71 for the PuBe source and 0.75 for the AmBe source with a relative error each of less than 15 %.

Zhenzhou Liu et al. [15] theoretically calculated the gamma to neutron ratio for a Chinese made AmBe neutron source for the 4.438 MeV gamma line as 0.57 ± 10 %. The reason the researchers limit themselves to the 4.438 MeV gamma line is that this energy represents the first excited state in the 9Be($\alpha$,n)12C reaction making it a prominent gamma emission line in AmBe sources. The discrepancy to G.Venkataraman et al.[22] obtained value is evident with the researchers explaining that most papers support a gamma to neutron ratio close to their findings.

The neutron emission energy spectrum and absolute neutron yield of a PoBe source has been determined in 1954 by Leona Stewart [20]. A low energy source gamma background has been mentioned along with an absolute neutron yield $1.2*10^6$ neutrons/sec of neutrons greater than 0.5 MeV for a source which consists of 13 g Po-210 and 7 g Be.

The researchers C. M. Cialella and J. A. Devanney [4] determined the PuBe gamma to neutron ratio to be 1.41.

The researchers D.M.Drake, J.C.Hopkins et al. [6] determined the PuBe gamma to neutron ratio to be 0.54.

# 3 Theory

## 3.1 Liquid Scintillator Detector

Liquid scintillator detectors contain a sealed scintillation liquid instead of a single crystal as typically used in NaI or LaBr detectors. The sealed liquid organic scintillator material contains a hydrocarbon liquid and is connected to a photomultiplier (PMT) via a photocathode. PMT signals are read out with a computer and are sampled digitally in this paper. The expected pulse signal for such a system is shown in Figure 2 [11]. The signal is depended on the scintillator detectors detection efficiency, scintillation efficiency, light collection efficiency and Quantum Efficiency [11]. For an impinging particle, the peak height of the signal is dependent on the fraction of the energy deposited in the scintillator.

For a gamma photon absorption, its most probable interactions with the scintillation material are the photoelectric effect, Compton effect and pair production. Most neutron detections originate from collisional energy transfer of fast neutrons [11]. Part of a Neutrons' kinetic energy is transferred to a detector atom nucleus, forming a recoil nucleus that is seen as an ionizing particle in the scintillator material.

Organic liquid scintillator detectors offer a fast response and decay time of typically tens of ns allowing for PSD and coincidence analysis. The shape of an organic liquid scintillator decay pulse can be described as the sum of two exponential decay functions of different decay constants [11]. Gammas and neutrons have different stopping powers in the organic liquid scintillator detector yielding different decay constants which result in different pulse decay signals. This difference in decay signal shape is exploited in PSD [11].

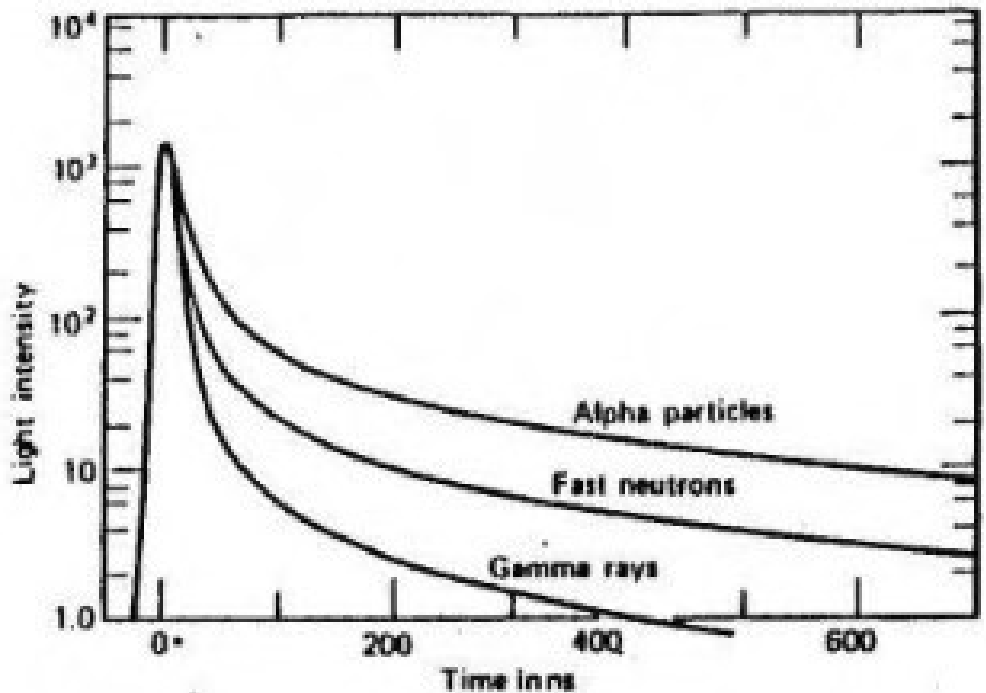

Figure 2: Effect of decay constants on the output light intensity in the liquid scintillator material. Credit: Knoll [11]

As seen in Figure 2, the pulse signal may not return to baseline due the decay constant. The output signal of a liquid organic scintillators system can exhibit baseline shift and baseline drift effects. The latter is typically caused by pulse tails in frontend electronics and by dark currents induced by the PMT and the photocathode [11].

## 3.2  CAEN Digital Pulse Processing

The CAEN Digital Pulse Processing digitizes incoming pulse signals coming from the photomultiplier. It does so by sampling the analog input pulse signal with a flash ADC. These digital samples are read by a field-programmable gate array and stored in a circular memory buffer. This memory buffer has a programmable size.

The PSD-DPP software read out offers two functions. First is the PSD mode in which the space of the waveform from each pulse is recorded. The digitized waveform is recorded in the form of pulse height for each time stamp. The recording interval is 96ns with a time interval of 1ns. The second more is the coincidence detection mode using two detector input channels. This mode is configured to a coincidence window of 5 trigger cycles where one trigger cycle is equal to 8 ns leading to a coincidence window of 40 ns. If two signals occur within the coincidence window, they are considered coincident and are recorded. Parameters recorded is the coincidence time stamp along with the time differences between the two detector channel triggers. The coincidence window is also know as the multiplicity window.

## 3.3  Neutron and Gamma Pulse Shape Discrimination

Liquid organic scintillators exhibit pulse light intensity decay constants which are dependent on the type of the particle. A pulse originating from heavy charged particles is likely to exhibit a different decay constants than neutrons or gamma rays. This property allows for discrimination between neutron and gamma detections in the liquid organic scintillator detector.

Figure 3 shows 20 pulse waveform shapes of the recorded PuBe neutron source used in this report. It is apparent that some pulses deposit more energy into the detector than others. It is also apparent that some pulses feature different decay constants than others.

Some pulsed may be read erroneously as a neutron particle or a gamma particle. If a neutron detection is considered to be the goal of the PSD technique, it inherently suffers from false positive and false negative detections. Typically a PSD measurement of a gamma only source such as Cs-137 source is coducted to determine the false positive rate. Using this method, Jennifer L. Dolan et al. [5] reported a false positive rate of 2 to 3 neutrons per 1 000 photon detection events. The researchers noted that false positive rates of coincidence false of multiple neutrons is expected to be near zero with false positives affecting primarily single neutron detection events.

An single organic liquid scintillator detector is used in this application of PSD. PSD maybe be conducted via the methods of the Tail-to-Full Integral Ratio, the Tail-to-Head Integral Ratio and the Tail-to-Pulse Height Ratio. Other PSD methods include Charge Integration Method (CIM), Pulse Gradient Analysis (PGA), Zero-Crossing Method (ZCM) and Discrete Wavelet Transform (DWT). In this paper, the Tail-to-Full Integral Ratio is used.

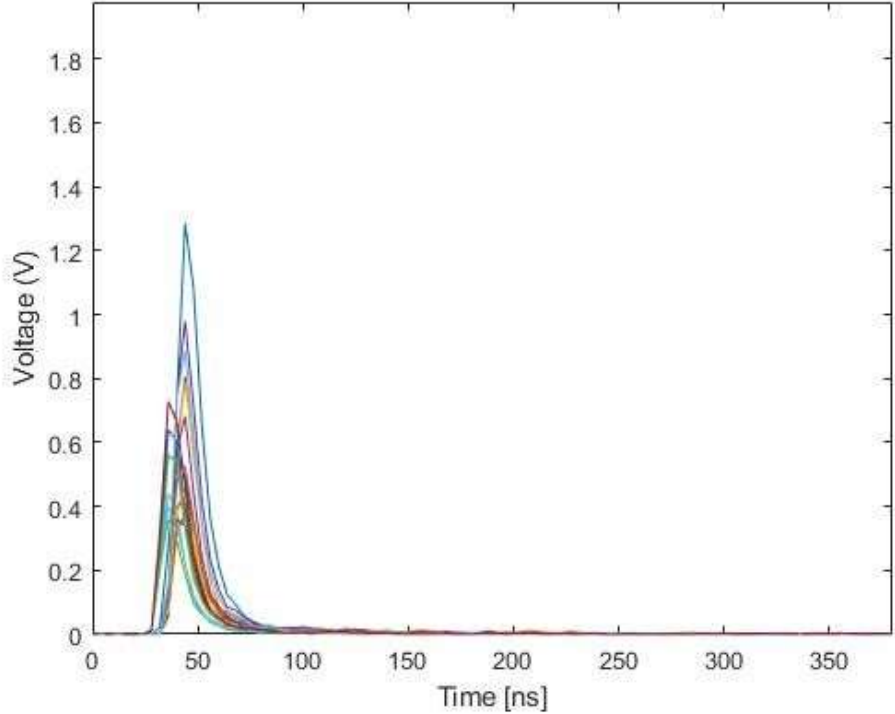

Figure 3: Plot of 20 PSD pulse waveforms from the PuBe neutron source utilized in this report.

Detailed explanation of the Tail-to-Full Integral Ratio is found in literature [11] [17]. The Tail-to-Full Integral Ratio PSD method conducts charge integration over each digitized signal pulse waveform to determine the so-called PSD parameter. The Tail-to-Full Integral Ratio PSD parameter is given by equation 1 [19].

$$PSDParameter = \frac{\int_{time_{tail}}^{time_{end}} Q(t)dt}{\int_{time_{begin}}^{time_{end}} Q(t)dt} \quad (1)$$

Whereas dt runs from 0ns to 96ns, Q(t) is the digitized pulse shape and *time_{begin}* is the time 0ns, the beginning of the pulse.

$time_{tail}$ is the time of the beginning of the pulse tail and $time_{end}$ is 96ns, the end of the pulse.

A figure of merit (FOM) can be defined for the Tail-to-Full Integral Ratio PSD method as the distance between the gamma ray and neutron peaks in units of the PSD parameter divided by the full width at half maximum (FWHM) of the peaks in units of the PSD parameter. A higher FOM is better.

Another class of PSD methods are those that exploit the differences between the frequency spectrum of gamma and neutron detection signals. An example is the frequency gradient analysis (FGA) PSD method. The Discrete Fourier Transform is used to discriminate between low and high frequency components in the pulse waveform signal.

### 3.4  Coincidence Measurement

A coincidence detection system can be described in a simple form to consist of two radiation detectors as a pulse source and a multi-channel time spectroscopy system. Each detector is attached to one time-pick-off device. One of the detectors is connected to a fixed delay. Each time-pick-off device is then connected to a time-to-amplitude converter (TAC) which sends an output pulse to a multi-channel analyzer (MCA) [11].

The TAC output signal is the pulse height. The pulse height in the TAC is directly proportional to the time difference between the two time-pick-off input pulses. This can be analyzed in a MCA resulting in a multi-channel time spectrum. Each channel number is proportional to a time differential between the TAC input pulses. The fixed delay mentioned above results in a time delay between the two pulses. This leads to a shift of a given pulse peak to higher channels.

The peak formed in the multi-channel time spectrum is known as the coincidence peak. The FWHM of the multi-channel time spectrum peak arises from statistical electrical processing errors such as time jitter and statistical errors arising from radiation scattering such as false coincidences from background radiation.

The TAC may be replaced with a coincidence unit. Only signals are selected whose time differentials is less than a preset time parameter. The MCA is replaced with a simple counter. This technique is known as the "Offline Method" and is be employed in this experiment. Digital waveform processing by CAEN is used to implement this detection method.

Another coincidence detection technique is known as the "Online Method" [11]. It involves summing up two individual pules of each detector. Time pick-off modules and the TAC are no longer employed. They are replaced with a signal summation module. The MCA is replaced with a Discriminator. The criteria of coincidence signal detection is that if the sum of the two pulses is greater than that of a preset lower level discrimination level, a coincidence occurred [11].

## 3.5  Neutron Sources

G.Venkataraman et al. [22] describes that relevant decay modes of the 13C* compound nucleus in the 9Be($\alpha$,n)12C reaction are given by the equations:

13C* → 12C + n
13C* → 12C* + n → 12C + n + $\gamma$(4.43MeV)
13C* → 12C* + n → 8Be + $\alpha$ + n → 3$\alpha$ + n

In (α,n) sources, the three major contributions to photons are the source background, α source gamma and z-ray emissions as well as 4.43 MeV gamma rays from the 12C* de-excitation.

The fission neutron spectrum of spontaneous fission in Cf-252 is well understood [12][11][13]. The neutron emission spectrum is comparable to that of an induced fission spectrum. Typically the average neutron energy is 2 MeV while the average neutron energy for (α,n) source range from 4 and 5 MeV due to a harder neutron spectrum. A typical (α,n) source features a neutron yield on the order of $10^6$ neutrons/sec per Ci compared to $10^9$ neutrons/sec per Ci for Cf-252 sources [16]. Fission neutrons are expected to be accompanied with prompt and delayed gamma emission along with prompt and delayed neutron emission.

## 3.6 Gamma to Neutron Ratio

The gamma to neutron ratio, denoted as R, is given by Zhenzhou Liu et al. [15] to be calculate by equation 2.

$$R = \int_0^{Emax} \chi(E_\alpha) \left[ \frac{\int_0^{E\alpha} \sigma_{\gamma_{12C*}}(E)/(dE/dx)dE}{\int_0^{E\alpha} \sigma_n(E)/(dE/dx)dE} \right] dE \quad (2)$$

$\chi(E_\alpha)$ is the fraction of α particles with energy $E_\alpha$. $\sigma_{\gamma_{12C*}}(E)$ is the cross section for the 9Be(α,n)12C* reaction with the 12C being in the first excited state at an energy of 4.43 MeV. $\sigma_n(E)$ is the cross section for all 9Be(α,n) reactions. $(dE/dx)$ is the stopping power of α particles in 9Be.

The theoretical gamma to neutron ratio R is not calculated in this paper due to the complexity of double integration of energy dependent cross sections.

| | |
|---|---|
| Neutron Intrinsic Detection Efficiency for PuBe | 34% |
| Gamma Intrinsic Detection Efficiency | 15% |
| Gamma to Neutron Ratio Efficiency correction fo PuBe | 44% |
| | |
| Neutron Intrinsic Detection Efficiency for PoBe | 30% |
| Gamma Intrinsic Detection Efficiency | 15% |
| Gamma to Neutron Ratio Efficiency correction for PoBe | 50% |

Figure 4: Estimated gamma to neutron ratio efficiency correction factors using data from F.Pino et al. [18].

## 4    Method

The goal of the conducted experiments is to quantify and compare coincidence detection rates of the PoBe neutron source, the neutron to gamma count ratio for the PoBe and PuBe neutron sources. A coincidence detection rate is determined experimentally for a PoBe source. For the other neutron sources coincidence detection rates are not evaluated due to COVID-19 related research limitations. The neutron to gamma count rate ratio is measured for both the PoBe and PuBe neutron sources. For the other neutron sources, the neutron to gamma count rate is taken from literature due to COVID-19 related research limitations.

## 5    Experimental Setup and Methods for PSD of PoBe Neutron Source

The goal of this experimental setup is to obtain the greatest possible absolute detection efficiency to determine the lowest possible gamma to neutron ratio for a PoBe source. This experiment is conducted on March 3rd, 2020, in the

Pennsylvania State University's Academic Projects Building, Laboratory Room 115.

A PoBe neutron source is utilized. The source consists of a commercially available Static Master 1U400 Self Powered $\alpha$ Ionizer of 1 inch in length and 1 inch in width. This is a Polonium210 source with an activity of 500 uCi at the time of manufacture. The Po-210 source is affixed in its housing by the manufacturer rendering it safe to handle without contamination risks. The source was manufactured by Amstat Industries in August 2019 and the experiment was conducted on March 3rd, 2020, resulting in a 215-day difference. Po-210 has a relatively short half-life of only 138.4 days, meaning that the source underwent roughly 1.5 half-lives. The Po-210 source activity is then calculated to be 170 uCi at the time of experimentation following the exponential decay law. A 0.1mm thick 99.99 % pure Beryllium sheet is placed on top of the Po-210 source to form a neutron source. An image of the Static Master 1U400 Self Powered $\alpha$ Ionizer is shown in Figure 5 and an image of the PoBe neutron source is shown in Figure 6.

The neutron source is placed in front of a cylindrical organic liquid scintillator manufactured by ELJEN with model number EJ-309. Its detection volume is 3" in diameter and 3" in length. The scintillator is connected to a Hamamatsu R1250 photomuliplier tube. The photomuliplier is attached to a CAEN NDT1471.

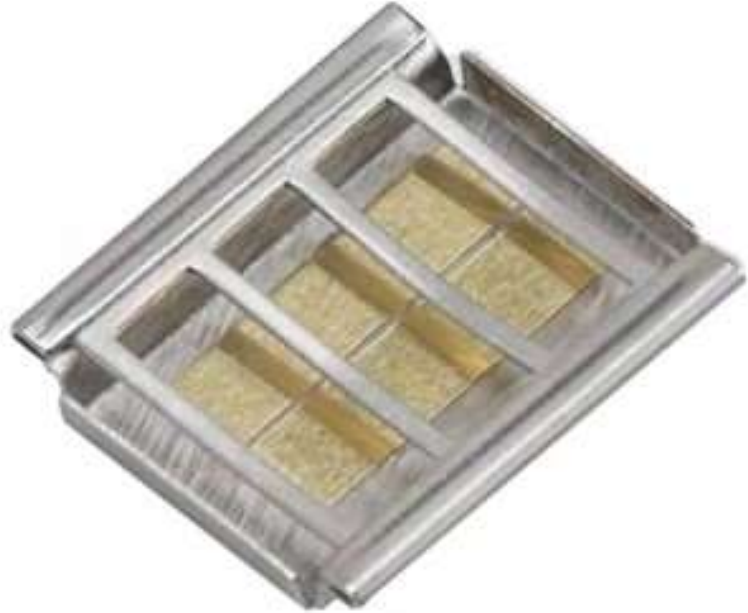

Figure 5: Manufacturer image of the Static Master 1U400 Self Powered $\alpha$ Ionizer. Credit: Amstat Industries

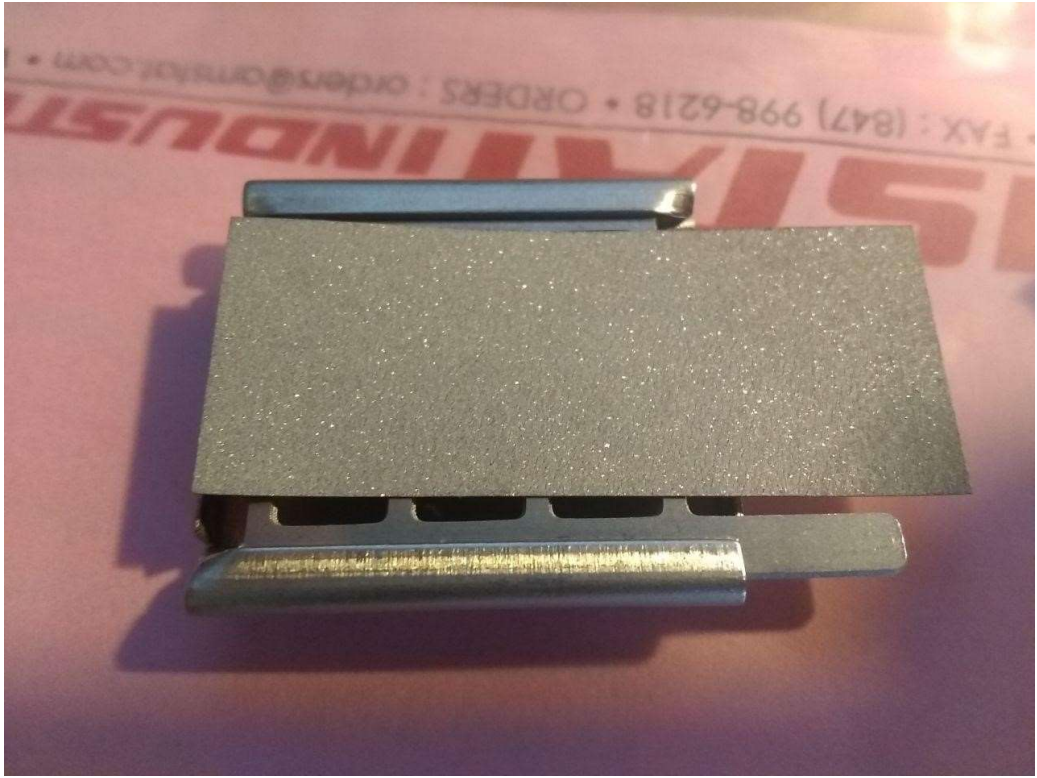

Figure 6: Image of the assembled Static Master 1U400 Self Powered ($\alpha$,n) neutron source.

High Voltage (HV) power supply using a RG-59B cable. Photomuliplier output signals are fed into a CAEN DT5720 signal digitizer by RG-316 cables. The CAEN DT5720 is controlled and read out by a Dell Optiplex 790 personal computer with the operating system Microsoft Windows 7 via a Universal Serial Bus (USB) connection. The CAEN DPP-PSD Control Software is used to control the CAEN DT5720 controller module.

The source is first placed 1 meter away from a cylindrical 3inch by 3inch organic liquid scintillator detector to record a background measurement. The closest distance is 1mm. The PoBe source is then moved closer to the detector recording an expected increase in gamma and neutron flux. The measurement time is set at 3 minutes to account for the expected low neutron source strength. Measurement time was timed with a manual mouse click and may vary by up to 1s. The reason for using manual mouse clicks to time the measurement is because the DPP-PSD software did not function with an automatic timer. Acquired PSD data is saved in a .dat file and is processed after data acquisition with a Matlab code provided by Scott Wandel at the Pennsylvania State University in May 2014. The PSD method used for this experiment is the Tail-to-Full Integral Ratio.

Measurement time was timed with a manual mouse click and may vary by up to 1s. The measurement time was attempted to be set 3 minutes for all source measurement samples and 2 minutes for background measurements.

# 6 Experimental Setup and Methods for Coincidence Counting of PoBe Neutron Source

PoBe neutron sources are described in literature to be ideal neutron interrogation sources due to their non-expected multiple neutron coincidences. This aspect evaluated in this experiment. The goal of this experiment is a non-detection of coincident neutrons induced by a PoBe neutron source.

The experimental setup for the coincidence counting of the PoBe Neutron Source is as follows. The CAEN V1751 VME signal digitizer received signals from two cylindrical organic liquid scintillator manufactured by ELJEN with model number EJ309. Its detection volume is 3" in diameter and 3" in length. The CAEN V6533M VME HV power supply provides power to the

Hamamatsu R1250 photomuliplier tubes, one in each detector. Each photomuliplier is attached to a CAEN V6533M High Voltage (HV) power supply using RG-59B cables. Photomuliplier output signals are fed into a a CAEN V1751 signal digitizer by RG-316 cables. Channel 0 of the CAEN V1751 signal digitizer relates to the detector located on the right hand side of the measurement system and Channel 1 relates to the detector located on the left hand side of the measurement system as it is seen in Figure 7. The CAEN V1718 VME controller module are powered by a CAEN VME 8010, which is a VME bus powered crate.

The PoBe source was placed horizontally in the setup seen in Figure 7 with the Beryllium film facing up against gravity to hold it in place. The minimum distance between the PoBe source and the detectors is 1mm to keep a safe minimum distance between the detectors and the source. This distance is denoted at the 0cm detector-source distance. The small detector source distance may introduce erroneous coincidences due to a number of events such as background photons travelling along both detector volumes or by Compton scattered photons in one detector which may scatter into the other detector. This small distance is chosen in order to increase the absolute detection efficiency of each detector given the low neutron and gamma production rates which are expected in a weak PoBe source.

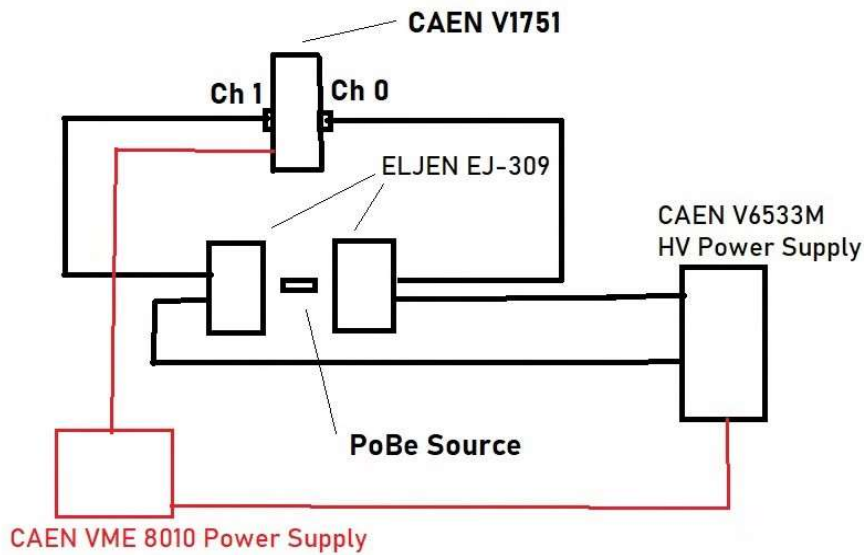

Figure 7: Schematic measurement setup for the PoBe coincidence measurement.

Coincidence counts are recorded for each detector. Pulse signals are digitized and are saved in a .dat file if they occur within the coincidence window. The coincidence window is set at 5 trigger cycles with each trigger cycle representing a 8ns time interval yielding a 40ns coincidence window. The .dat file saves a time stamp and a coincidence time shift between the detectors for each coincidence event. The data is analyzed post data acquisition. The number of coincidence events is simply determined by counting the number of recorded coincidence events found in the .dat file.

The coincidence rate of a PuBe source has three major contributions. First, the background is expected to contribute to coincidences due to the close arrangement of the two detectors. Second, the Po source itself is expected to contribute to the coincidence rate due to ($\alpha$,2x) reactions in its matrix, whereas x is any subatomic particle detectable by the organic liquid scintillators such as photons, electrons or protons. Third, once the PuBe source is assembled, neutrons and gammas are expected to arise leading to a PuBe coincidence measurement which has contributions from the background plus the Po Source plus the PoBe induced neutrons and gammas.

Three coincidence measurements are conducted each with 100s duration. First the background coincidence rate is measured. Then the Po source without the Be foil is measured. Then the PoBe neutron is measured. The number PoBe neutron and gamma detection events which are induced by the 9Be($\alpha$,n)12C* reaction are then calculated by: Induced PoBe neutrons and gammas = PoBe Source - (Po Source - Background) - Background. Error propagation is conducted to obtain a one sigma significance to this value.

## 7 Experimental Setup and Methods for PSD PuBe Neutron Source

The goal of this experimental setup is to obtain the greatest possible absolute detection efficiency to determine the lowest possible gamma to neutron ratio for a PuBe source. This experiment is conducted on March 3rd, 2020, in the Pennsylvania State University's nuclear fuel building. The data for this experiment was taken from a previous report by Sebastian Ritter and Florian Passelaigue issued on March 11th 2020 and titled" Neutron and Gamma Pulse Shape Discrimination using a Liquid Scintillator Detector" due to COVID-19 related research restrictions.

   The source is initially placed 100 inches away from a cylindrical 3inch by 3inch organic liquid scintillator detector to record a background measurement. The PuBe source is then moved closer to the detector recording an expected increase in gamma and neutron flux as the absolute detection efficiency increases. The lowest measured gamma to neutron ratio will be determined and is expected to arise at the closest PuBe source to detector distance. The measurement time is timed with a manual mouse click and may vary by up to 1s. The measurement time is attempted to be set at 5s for all samples.

The neutron source is placed in front of a cylindrical organic liquid scintillator manufactured by ELJEN with model number EJ-309. Its detection volume is 3" in diameter and 3" in length. The scintillator is connected to a Hamamatsu R1250 photomuliplier tube. The photomuliplier is attached to a CAEN NDT1471 High Voltage (HV) power supply using a RG-59B cable. Photomuliplier output signals are fed into a CAEN DT5720 signal digitizer by RG-316 cables. The CAEN DT5720 is controlled and read out by a Dell Optiplex 790 personal computer with the operating system Microsoft Windows 7 via a Universal Serial Bus (USB) connection. The CAEN DPP-PSD Control Software is used to control the CAEN DT5720 controller module.

The PuBe source constitutes a central PuBe source embedded into a polymer matrix for moderation. The whole PuBe assembly is placed into a steel drum for regulatory compliance and ease of use. The PuBe source consists of 1Ci of Pu-239 that has been integrated and produced on August 27th, 2014. Pu-239 has a half-life of roughly 24110 years and thus its decay from its initial activity can be neglected.

An image of the experimental setup is shown in Figure 8.

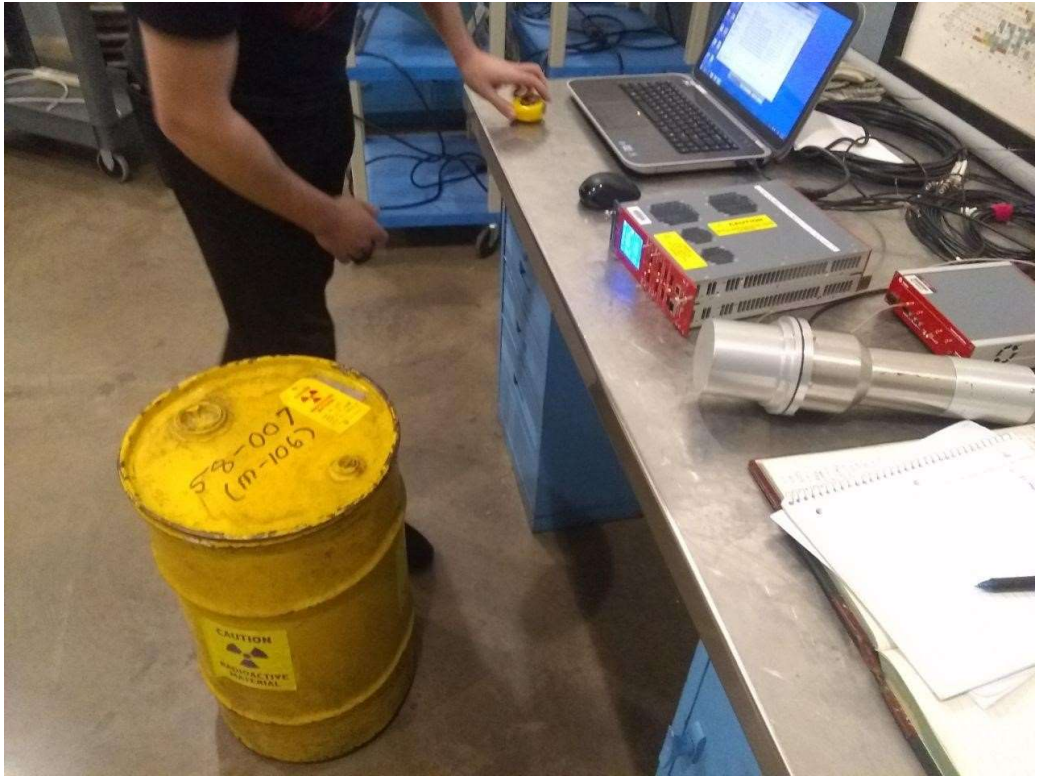

Figure 8: An image of the experimental PuBe PSD measurement setup.

## 8 Results and Discussion

The measured gamma to neutron ratios as well as gamma to neutron ratios found in literature are plotted in figure 13. The gamma to neutron ratios of the background radiation in the laboratory was found to be 455.1666667 ± 558.0750771 for a 3-minute measurement and in the Pennsylvania State University's nuclear fuels building 27.81818182 ± 6.036515482 for a 5 second measurement. The gamma to neutron as a function of distance is determined for both the PuBe and PoBe source.

The gamma to neutron ratio as a function of distance for the PuBe source is shown in figure 11. Figure 16 shows the PSD plot of the PuBe neutron source after a measurement time of 5 seconds. Figure 17 shows the PSD plot of the PuBe neutron source background measurement with a measurement time of 5 seconds.

All measurements are made for 5 seconds. The PuBe source is measured in the nuclear fuels building. The ratio increases with distance as less source neutrons reach the detector and as the gamma background remains relatively stable. This figure also illustrates the strong dependency of the measured gamma to neutron ratio on the absolute detector efficiency. The geometric neutron detection efficiency decreases with increasing detectorsource distance leading to an increase in the gamma to neutron ratio. The last measurement point in figure 11 is regarded as a background measurement. The nuclear fuel building has a high neutron flux environment with a gamma to neutron ratio of 27.81818182 ± 6.036515482 for a 5 second measurement. Table 3 shows the acquired data set for the PuBe neutron source with a measurement time of 5 seconds. The gamma to neutron ratio for the PuBe source is determined to be 8.34 ± 0.33.

Table 1: Acquired data set of the PoBe neutron source coincidence measurement. Measurement time is 100s.

| Three Coincidence Measurements with 100s each | | | |
|---|---|---|---|
|  | counts | one sigma counts | relative one sigma uncertainty |
| Background | 213.33 | 10.56 | 4.95% |
| Po Source + Background | 239.57 | 8.94 | 3.73% |
| PoBe Source + Po Source + background | 271.12 | 9.51 | 3.51% |
|  |  |  |  |
| PoBe source | 31.55 | 19.83 | 62.87% |

Table 2: Acquired data set of the background measurement for the PoBe neutron source. Measurement time is 2 minutes.

| Background PSD 2min | | | | |
|---|---|---|---|---|
| Experiment Run | Neutron Count | One Sigma Neutron Count | Gamma Count | One Sigma Gamma Count |
| No 1 | 1 | 1 | 923 | 30.38 |
| No 2 | 3 | 1.73 | 951 | 30.84 |
| No 3 | 2 | 1.41 | 857 | 29.27 |
| | | | | |
| Average | 2 | 2.45 | 910.33 | 52.26 |

Table 3: Acquired data set for the background measurement and for the 4 inches source-detector distance measurement for the PuBe neutron source. Measurement time is 5 seconds for both the background and the PuBe neutron source.

| PuBe Neutron Source PSD 5s | | | | |
|---|---|---|---|---|
| | Neutron Count | One Sigma Neutron Count | Gamma Count | One Sigma Gamma Count |
| 4 inches Detector-Source Distance | 715 | 26.74 | 5963 | 77.22 |
| Background | 22 | 4.69 | 612 | 24.74 |

The gamma to neutron ratio as a function of distance for the PoBe source is shown in Figure 12. Each measurement is conducted for the duration of 3 minutes. Figure 10 shows the acquired data set. The gamma to neutron ratio for the PoBe source is determined to be 2.26 ± 0.09. Figure 14 shows the PSD plot of the PoBe neutron source after a measurement time of 2 minutes.

The PoBe measured is in the laboratory room. The ratio strongly monotonically increases with distance as less source neutrons reach the detector and as the gamma background remains relatively stable. The laboratory room has a low neutron flux environment with a gamma to neutron ratio of 455.1666667 ± 558.0750771 for a 2-minute measurement. Table 2 shows the acquired data set of the background measurement for the PoBe neutron source. Figure 15 shows the PSD plot of the static master 1U400 alpha source after a measurement time of 2 minutes.

The gamma to neutron ratios of the two neutron sources are found to hold no simple mathematical relationships with detector-neutron source distance besides a monotonic increase. This is attributed to the complex neutron source and detector dimensions along back neutrons scattering back into the detector from inelastic and elastic scatterings in air and surrounding materials.

The high PoBe and PuBe gamma to neutron ratio is attributed to the fact that the intrinsic gamma and neutron detection efficiencies have not been properly calculated and are merely estimate by a paper from F.Pino et al. [18]. Furthermore, it is shown in Figures 11 and 12 that an increase in distance between detector and neutron source yields significantly higher gamma to neutron ratios due to decreasing geometric neutron detection efficiencies. Increasing the geometric detection efficiency to 4pi and moving the 4pi detector so close to the source that neutrons are unlikely to interact in the medium between source and detector could be expected to decrease the gamma to neutron ratio from experimental results in this paper.

When conducting PSD measurements with the same detector, both source gamma-rays and neutrons are subject to the same geometrical factors that contribute to the absolute detection efficiency. Only the relative intrinsic detection efficiencies between gamma-rays and neutrons of the detector may lead to correction factor for the measured gamma to neutron ratio. F.Pino et al. [18] noted that detection efficiencies of neutrons and gamma rays of the EJ-309 organic liquid scintillator detector depends strongly on the size of the detector. The researchers used a 51mm-by-51mm cylindrical detector. This size is roughly comparable to the one utilized in this experiment of 3 by 3 inches in size. From their best fits of their measurements with MCNPX calculated data this report takes 34 % detector efficiency for PuBe neutrons with an average neutron energy is 2 MeV and 30 % for ($\alpha$,n) neutron sources which range in average energies from 4 and 5 MeV. 4.43 MeV gamma rays are expected to arise from the 12C* de-excitation. Other low energy gamma rays may be present in the detector volume. 1.5 MeV is assumed for the average gamma ray energy yielding roughly 15 % detector efficiency for gamma rays according to the paper by F.Pino et al. [18]. The estimated gamma to neutron ratio efficiency correction for the PuBe and PoBe neutron sources is shown in Figure 4. A more thoroughly analysis of absolute gamma and neutron detection efficiencies is recommended for future reports.

Using the estimated gamma to neutron ratio efficiency correc-tion factors shown in Figure 4 the obtained gamma to neutron ratios as well as the gamma to neutron ratios found in literature are plotted in Figure 9. The gamma to neutron source ratio is found to be 3.67 ± 0.14 and 1.13 ± 0.05 for the PuBe and for the PoBe source respectively.

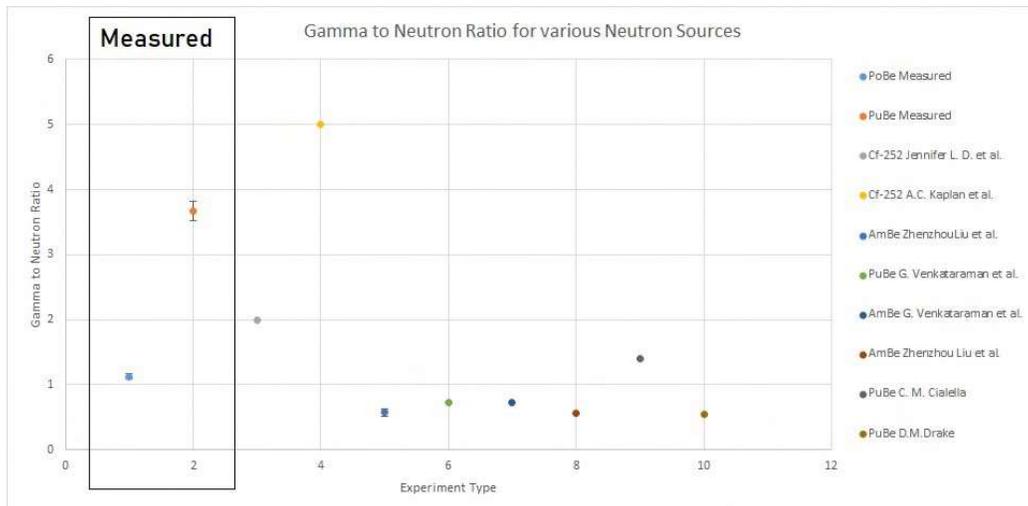

Figure 9: Corrected gamma to neutron ratios from experimental results for the PuBe and the PoBe neutron sources are shown in the boxed area. These gamma to neutron ratios have been corrected for different intrinsic detection efficiencies of neutrons and gamma-rays in the EJ-309 organic liquid scintillator following results from F.Pino et al. [18]. Also theoretical and experimental gamma to neutron ratios from litrature are shown for AmBe, PuBe and Cf-252 neutron sources.

The coincidence measurements are conducted on the PuBe source. The resulting data is shown in Figure 18. It is evident that PoBe induced neutrons and gamma detections amount to 31.55 ± 19.83 or 62.87 %. The detection window time is 100s. This data could be interpreted as being inconclusive and one cannot rule that a successful non-detection has occurred. However, considering that the induced coincidence rate is significantly below the natural background with a large 63 % uncertainty, one can also conclude that a non-detection is successful.

| | PoBe Neutron Source PSD 3min | | | | | | | |
|---|---|---|---|---|---|---|---|---|
| | Distance [cm] | | | | | | | |
| | 0 | 0.5 | 1 | 2 | 5 | 10 | 20 | 40 |
| Neutron Count | 820.00 | 324.00 | 289.00 | 196.00 | 133.00 | 57.00 | 24.00 | 4.00 |
| Neutron Count Sigma | 28.64 | 18.00 | 17.00 | 14.00 | 11.53 | 7.55 | 4.90 | 2.00 |
| Gamma Count | 1855.00 | 1557.00 | 1550.00 | 1429.00 | 1447.00 | 1422.00 | 1408.00 | 1369.00 |
| Gamma Count Sigma | 43.07 | 39.46 | 39.37 | 37.80 | 38.04 | 37.71 | 37.52 | 37.00 |
| | | | | | | | | |
| Gamma to Neutron Ratio for PuBe | 2.26 | 4.81 | 5.36 | 7.29 | 10.88 | 24.95 | 58.67 | 342.25 |
| Gamma to Neutron Ratio Sigma | 0.09 | 0.29 | 0.34 | 0.56 | 0.99 | 3.37 | 12.08 | 171.37 |

Figure 10: The acquired data set for PoBe PSD measurements. The gamma to neutron ratio for the PoBe source is determined to be 2.26 ± 0.09 without correcting for different intrinsic detection efficiencies of neutrons and gammarays in the EJ-309 organic liquid scintillator.

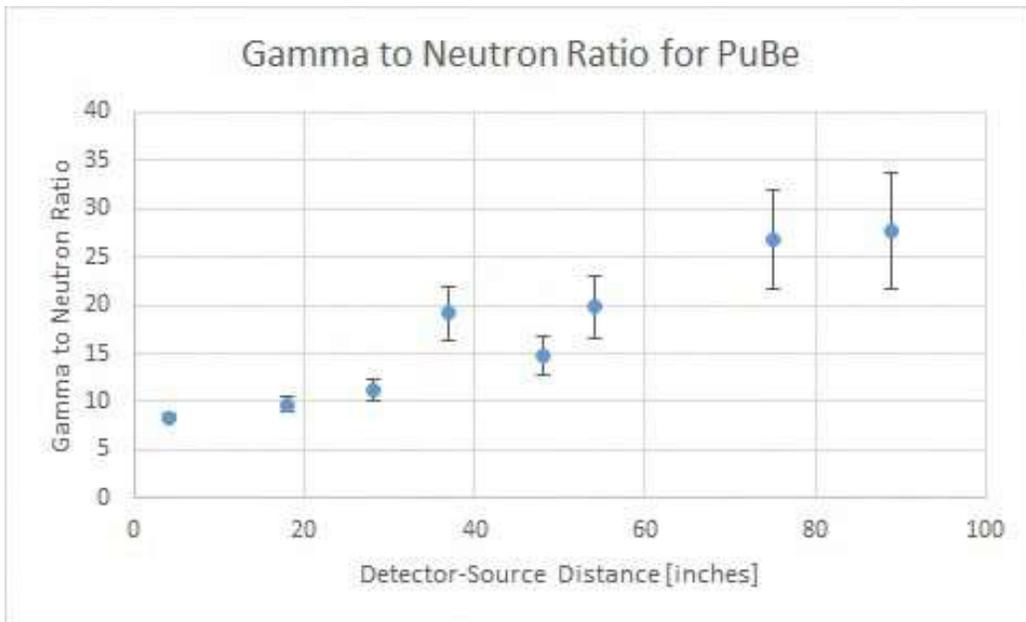

Figure 11: The gamma to neutron ratio as a function of distance for the PuBe source. An increase in distance between detector and neutron source yields significantly higher gamma to neutron ratios due to decreasing geometric neutron detection efficiencies.

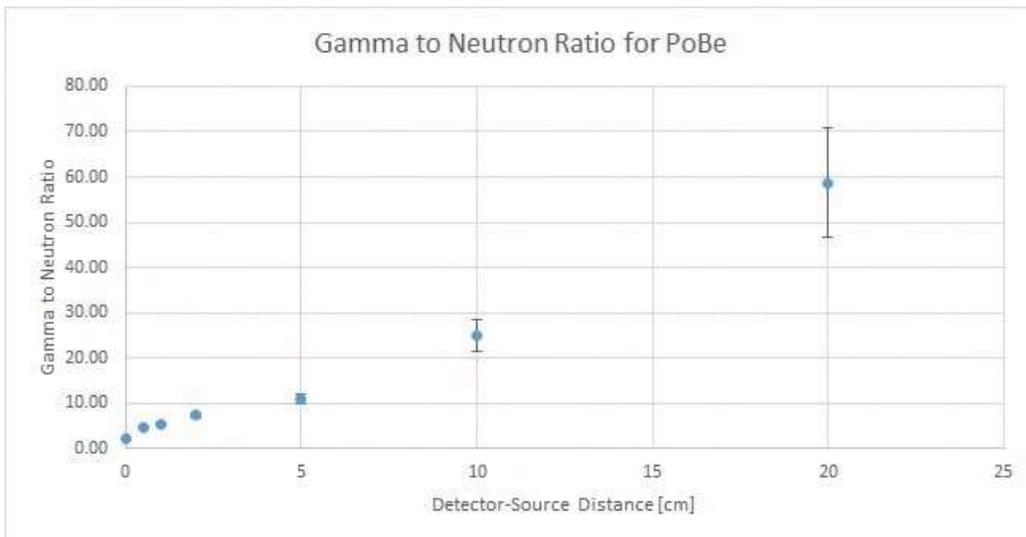

Figure 12: The gamma to neutron ratio as a function of distance for the PoBe source. As it is suspected with the PuBe source, an increase in distance

between detector and neutron source yields significantly higher gamma to neutron ratios due to decreasing geometric neutron detection efficiencies.

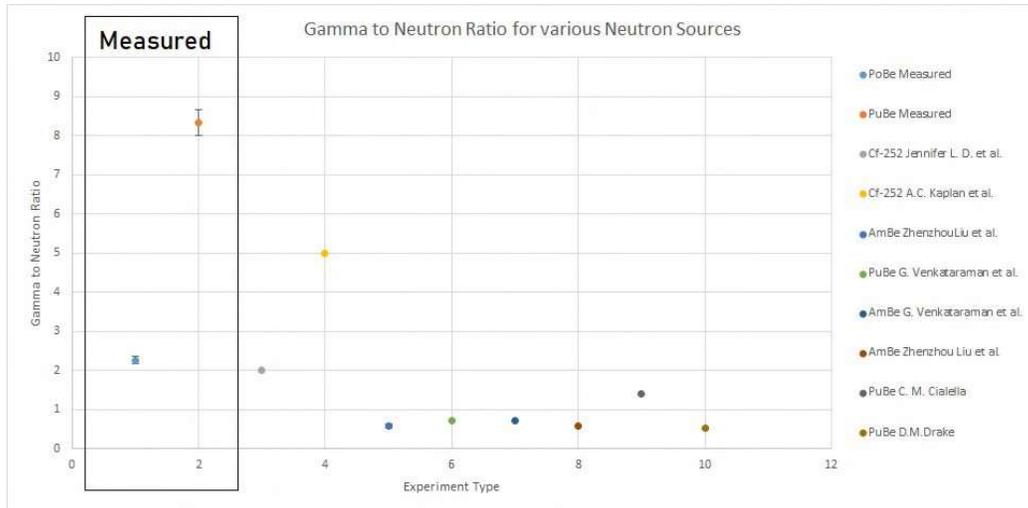

Figure 13: Uncorrected gamma to neutron ratios from experimental results for the PuBe and the PoBe neutron sources are shown in the boxed area. Also theoretical and experimental gamma to neutron ratios from litrature are shown for AmBe, PuBe and Cf-252 neutron sources.

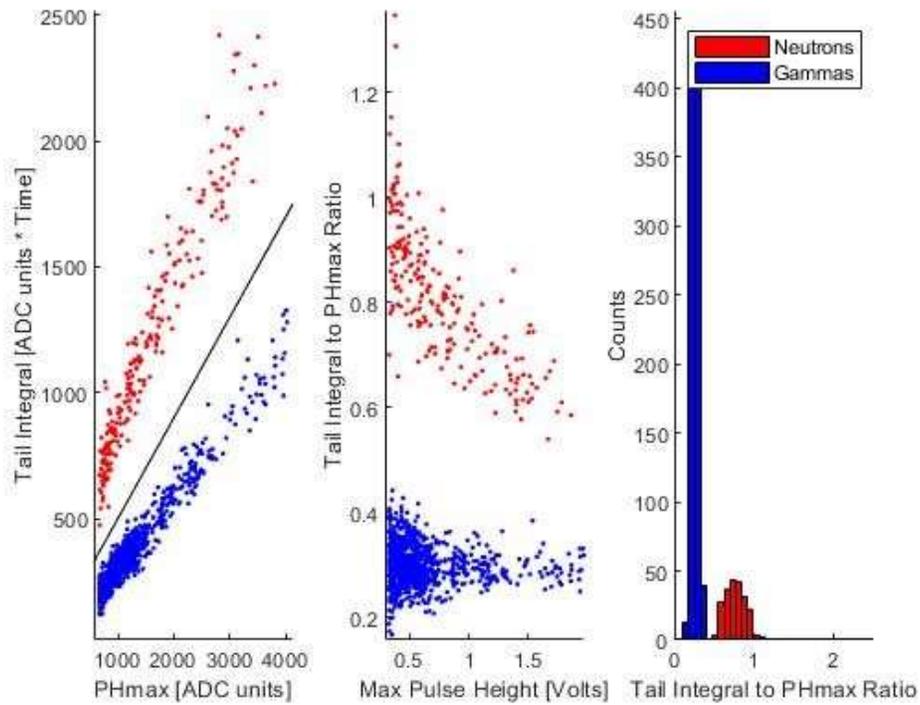

Figure 14: PSD plot of the PoBe neutron source after a measurement time of 2 minutes.

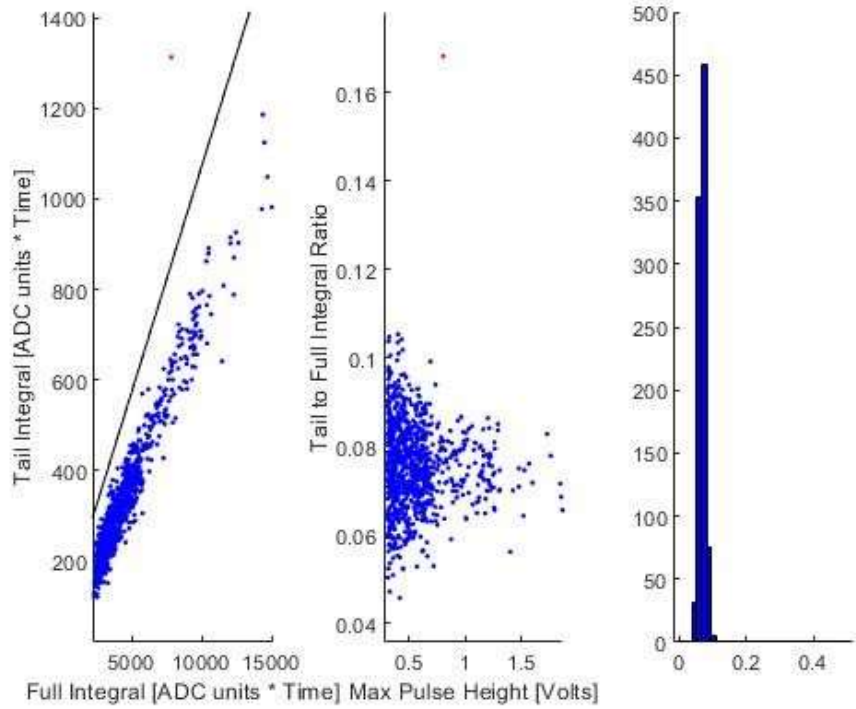

Figure 15: PSD plot of the static master 1U400 alpha source after a measurement time of 2 minutes.

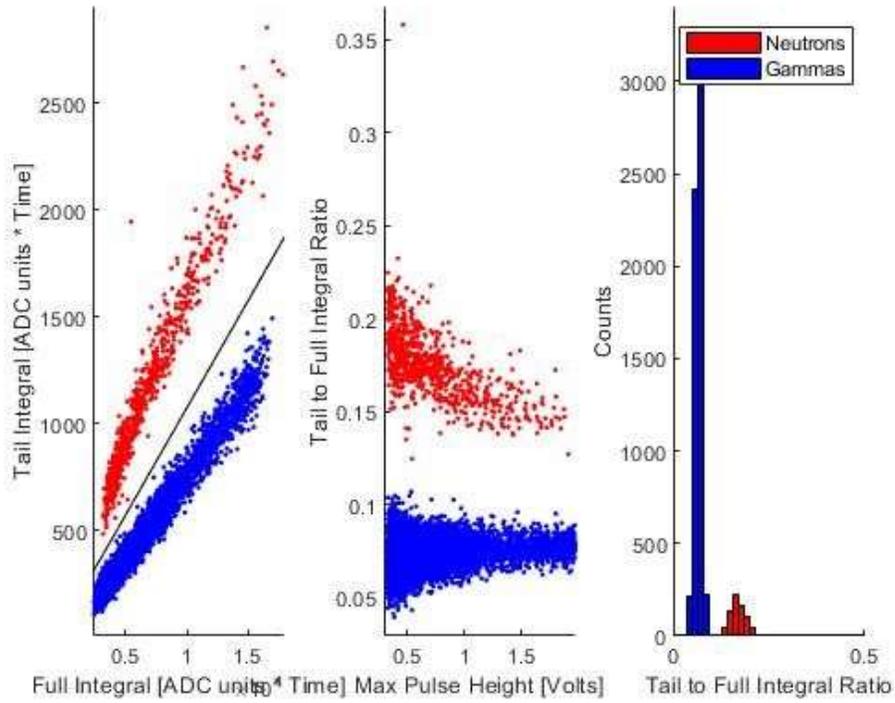

Figure 16: PSD plot of the PuBe neutron source after a measurement time of 5 seconds.

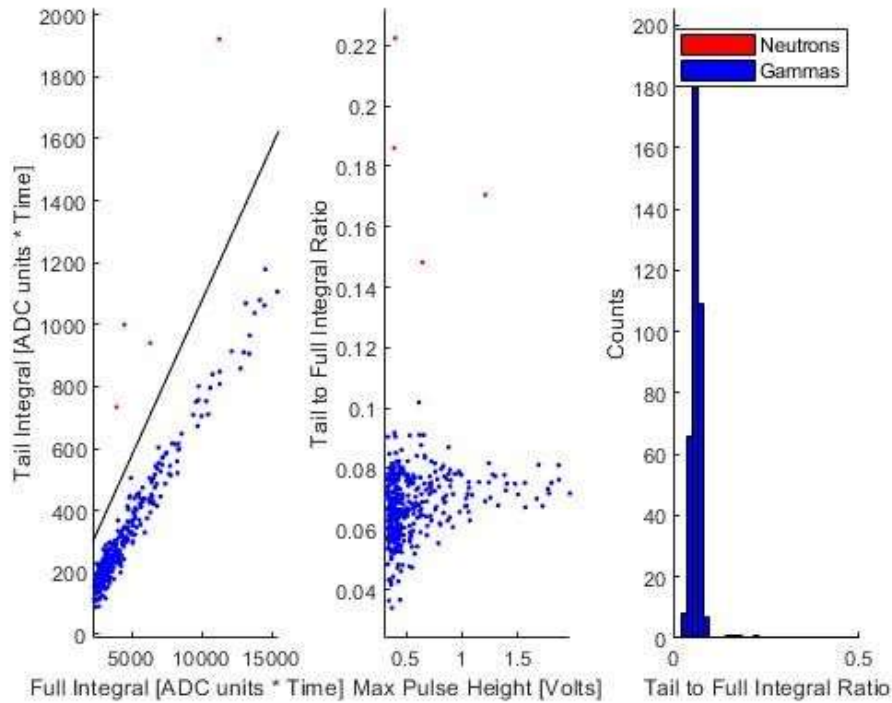

Figure 17: PSD plot of the background environment (the Pennsylvania State University's nuclear fuels building) in which the PuBe neutron source is measured. Measurement time is 5 seconds.

| 3 Coincidence Measurements with 100s each | | | |
|---|---|---|---|
| | counts | one sigma counts | relative one sigma uncertainty |
| Backgound | 213.33 | 10.56 | 4.95% |
| Po Source + Background | 239.57 | 8.94 | 3.73% |
| PoBe Source + Po Source + background | 271.12 | 9.51 | 3.51% |
| | | | |
| PoBe source | 31.55 | 19.83 | 62.87% |

Figure 18: Acquired data set of coincidence measurements of the PuBe source.

Table 1 shows the acquired data set of the PoBe neutron source coincidence measurement. Measurement time is 100s.

The confidence level of non-detection may also be calculated using the T-Test. If we consider the Po source to be the background and PoBe to be the source, one can calcualte a T-value as ($Xe - Xb$)/$SQRT(Xe + Xb)$. In this case, Xe equals to 271.12 counts and Xb equals to 239.57 counts. The T-Value equals 1.396 and a one-sided test is conducted. The confidence level is 91.77 % that a coincidence detection has occurred for the PuBe source. This result implies that the applied neutron source will produce two or more coincident gamma-rays or neutrons with a confidence level of 91.77 %.

Literature states that ($\alpha$,n) sources do not produce coincident neutrons. Coincident gamma-rays may still be possible. Since the measurement included both neutrons and gamma-rays one cannot conclude that coincident neutrons were measured with a confidence level of 91.77 %. The only cnclusion that can be drawn is that the applied PoBe neutron source will produce two or more coincident gamma-rays or neutrons with a confidence level of 91.77 %. Thus, the result is not in conflict with literature.

# 9 Conclusion

A comparative neutron source analysis is conducted on a PuBe source and a PoBe source using organic liquid scintillator detectors. The PuBe source is a rudimentary constructed source from commercially available products. Parameters of neutron source classification are the neutron to gamma count rate ratio. The ratios of both sources are found to be above literature values. Furthermore, the the confidence level of detection of coincident neutron or gamma-rays is measured for the PoBe source. Methods used in this analysis are two organic liquid scintillator detectors and coincidence analysis using digital signal processing.

The result shows a confidence level of 91.77 %.